\documentclass[aps,pra,twocolumn]{revtex4-2}
\usepackage{amsfonts}
\usepackage{amsmath}
\usepackage{mathrsfs}
\usepackage{amssymb}
\usepackage{graphicx}
\usepackage{bm}
\usepackage{color}

\usepackage[colorlinks=true,linkcolor=blue,anchorcolor=blue,citecolor=blue,urlcolor=blue]{hyperref}

\begin{document}
	
	\title{Instabilities of a Bose-Einstein condensate with mixed nonlinear and linear lattices}
	
	\author{Jun Hong}
	\affiliation{Department of Physics, Shanghai University, Shanghai 200444, China}
	
	\author{Chenhui Wang}
	\affiliation{Department of Physics, Shanghai University, Shanghai 200444, China}
	
	\author{Yongping Zhang}
	\email{yongping11@t.shu.edu.cn}
	\affiliation{Department of Physics, Shanghai University, Shanghai 200444, China}

	\begin{abstract}
		
	Bose-Einstein condensates (BECs) in periodic potentials generate interesting physics on the instabilities of Bloch states. 	The lowest-energy Bloch states of BECs in pure nonlinear lattices are dynamically and Landau unstable,  which breaks down BEC superfluidity. In this paper we propose to use an out-of-phase linear lattice to stabilize them. The stabilization mechanism is revealed by the averaged interaction. We further incorporate a constant interaction into BECs with mixed nonlinear and linear lattices, and reveal its effect on the instabilities of Bloch states in the lowest band.

	\end{abstract}
	
	\maketitle
	
	\section{Introduction}
	
Since the first experimental realizations in 1995~\cite{Wieman1995,Davis1995,Bradley1995}, atomic Bose-Einstein condensates (BECs) have been fundamental platforms to explore quantum many-body phenomena. Theoretically, the dynamics of a BEC can be described well using the mean-field Gross-Pitaevskii (GP) equation~\cite{Pitaevskii2016}. A linearization of the GP equation around a BEC state becomes the so-called Bogoliubov-de Gennes (BdG) equation which describes elementary excitations of the corresponding BEC state~\cite{Castin2001}.  The distinguishing feature of the BdG equation is that the BdG Hamiltonian is non-Hermitian, which allows for the existence of  complex excitations.  In the presence of any complex modes, a small deviation from the BEC state may diverge exponentially with time, which destroys the BEC state. Such breakdown of the corresponding BEC state is referred to as the dynamical instability.  However, not all BEC states possess the dynamical instability.  A standard homogeneous BEC is dynamically stable and its elementary excitation is the gapless phonon mode in the long wavelength limit~\cite{Ozeri2005}.  The identification of the dynamical instability of  a BEC state becomes a fundamental issue. 
	
One of outstanding systems that nurture  dynamical instability is  BECs in optical lattices~\cite{Wu2001,Smerzi2002,Konotop2002}.  The optical lattices modify the dispersion relation of a BEC to give rise to Bloch spectrum.  The associated BEC Bloch states may be dynamically unstable due to the interplay between their dispersion and atomic interactions~\cite{Wu2001,Smerzi2002,Konotop2002}.  The dynamical instability of BEC Bloch states has been experimentally observed by measuring the decay of condensed atoms~\cite{Fallani2004}. The optical-lattice-induced dynamical instability relating to the breakdown of BEC superfluidity has been extensively studied~\cite{Mueller2002,Wu2003, Machholm2003}.  It can be approached analytically in a Kronig-Penney potential~\cite{Danshita2007}.   Attractive interactions~\cite{Barontini2007} or complicated atomic interactions~\cite{Xue2007} give the instability more features. Two dimensional optical lattices~\cite{Chen2010} and Bloch states in higher Bloch bands~\cite{Xu2013} are investigated theoretically.  Besides the dynamical instability, optical lattices can introduce Landau instability to BEC Bloch states~\cite{Wu2001}. The Landau instability happens when the BEC Bloch states are energetically unfavorable~\cite{Wu2003}. Generalization of a single BEC in optical lattices to multiple components attracts much attention~\cite{	Jin2005,Hooley2007,Liu2007,Ruostekoski2007,Baizakov2009,Barontini2009,Huang2010,
Hui2011, Penna2017,He2021,Yamashika2021}.  Multiple-component BECs in optical lattices combine rich phases due to numerous parameters and the optical-lattice-modified dispersion relation.  The dynamical instability of these systems presents complex features and becomes more interesting~\cite{Liu2007,Penna2017}.  Meanwhile, in the tight-binding regime of the optical lattices, the dynamical instability of multiple components is tractable in an analytical way~\cite{Jin2005,Ruostekoski2007,Baizakov2009,Huang2010}. Furthermore, the optical lattices can be component-dependent, which introduces rich instability structures~\cite{Hooley2007,Barontini2009,He2021}. In multiple-component BECs, one can define spin currents. Multiple-component BECs in optical lattices provide an important platform to study the dynamical instability of spin currents~\cite{Hui2011,Yamashika2021}.

Being different from BECs in optical lattices where interactions come from condensed atoms and optical lattices provide linear periodic potentials are nonlinear lattices, which represent spatially periodic modulation of the interatomic interactions~\cite{Torner2011,Watanabe2016}. Nonlinear lattices can be experimentally implemented by the controllable optical Feshbach resonances~\cite{Theis2004,Yamazaki2010,Clark2015}. The lowest-energy Bloch states (at Brillouin zone center in the lowest band) in nonlinear lattices are always dynamically unstable so that they can not support superfluidity~\cite{Shaoliang2013}. Only Bloch states at a finite quasimomentum regime close to Brillouin zone edge are dynamically stable~\cite{Shaoliang2013}. Physically, for these Bloch states, atoms are mainly confined into the negative parts of nonlinear lattices, and the empty-occupation of the positive parts behaviors like barriers to prevent the tunneling between the negative-part occupations, which stabilizes the corresponding Bloch states~\cite{Dasgupta2016}.    In most lattice experiments, BECs are usually prepared in the lowest-energy Bloch states. However,  the instability breaks the preparations in nonlinear lattices. Therefore, stabilizing the lowest-energy Bloch states becomes an important aspect for experimental realizations. Refs.~\cite{Shaoliang2013, Dasgupta2016,Xie2021} propose to add a constant interaction to nonlinear lattices for their stabilization. Nonlinear lattices with a constant repulsive interaction generate repulsive effective interatomic interactions providing a possible means for the stabilization.  Ref.~\cite{Watanabe2018} suggests that a linear and coherent Rabi-coupling between two-component BECs can be used to stabilize the lowest-energy Bloch states. 

In the present paper, we systematically study the instabilities of BECs with mixed nonlinear and linear lattices.  Nonlinear phenomena in mixed nonlinear and linear lattices have been widely studied~\cite{Torner2011}.  The interplay between nonlinear and linear lattices gives new properties to bright solitons~\cite{Abdullaev2007,Kartashov2008,Jiefang2010,Yaroslav2016,Shi2020} and increases their mobility~\cite{ Rapti2007}. These systems provide a possibility to study the effect of commensurability between two lattices on the existence of solitons~\cite{Sakaguchi2010}. The coexistence of two lattices helps to stabilize  solitary waves against collapse~\cite{Luz2010,Dong2011,He2012,Holmes2013}.  The spatially localized states in multiple-component BECs with mixed nonlinear and linear lattices are revealed to have interesting properties~\cite{Sekh2012,Xu2021}. More importantly, the mixed lattices are proposed to support long-time Bloch oscillations~\cite{Salerno2008}. So far, all studies on the mixed nonlinear and linear lattices are relevant to the existence and stability and dynamical management of solitary waves. Here, we study the instabilities of spatially extended waves, i.e., Bloch states, in mixed nonlinear and linear lattices. Bloch states are relevant to experimentally load BECs into the mixed lattices.  The instabilities of Bloch states in these systems relate to the  breakdown of BEC superfluidity. Therefore, our study is  experimentally involved.  We examine the dynamical instability and Landau instability of Bloch states in the lowest Bloch band in mixed nonlinear and linear lattices by analyzing the BdG equation. In comparison with the instabilities of BECs with pure nonlinear lattices as studied in~\cite{Shaoliang2013, Dasgupta2016}, we find that an out-of-phase linear lattice can assist to stabilize the Bloch states around the Brillouin zone center. We present the mechanism of the stabilization using the concept of the averaged interaction.  According to the mechanism, an in-phase linear lattice is useless for the stabilization.  We further reveal that the out-of-phase and in-phase linear lattices can modify the dynamical instability of the Bloch states around Brillouin zone edges; the out-of-phase linear lattice destabilizes the states and the in-phase lattice strengthens their stabilities. Meanwhile, we incorporate a constant interaction into the BECs with mixed lattices and study the effect of repulsive and attractive constant interactions on the instabilities of Bloch states.  

The paper is organized as follows. In Sec.~\ref{Model}, we present the theoretical framework for the study on the instabilities of BEC Bloch states in mixed nonlinear and linear lattices. It includes the GP equation and the derivations of the BdG equation.  In Sec.~\ref{Bands}, we show nonlinear Bloch spectrum and indicate the existence of nonlinear Bloch states.  The properties of Bloch states are shown by their density distributions.  In Sec.~\ref{Instability}, the dynamical and Landau instabilities of Bloch states are presented, with a purpose that a linear lattice can stabilize the Bloch states around the Brillouin zone center. The mechanism of the stabilization is uncovered. We also study the effect of a constant interaction. The conclusion follows in Sec.~\ref{Conclusion}.

	\section{Model}
	\label{Model}

We consider a BEC with spatially periodic modulated interactions in the presence of a linear optical lattice.  The system is described by the Gross-Pitaevskii (GP) equation as follows, 
\begin{equation}
	\label{gp}
	i\frac{\partial \psi}{\partial t}=-\frac{1}{2}\frac{\partial^{2} \psi}{\partial x^{2}}-V\cos (x) \psi+\mathcal{G}_\textrm{non}|\psi|^{2}\psi.
\end{equation}
$\psi(x,t)$ is the wave function of the BEC. 
The GP equation is dimensionless. We have already used the units of energy and length as 
 $8E_\textrm{rec}$ and
 $1/2k_{l}$ respectively. Here, the recoil energy of the optical lattice lasers is $E_\textrm{rec}=\hbar^{2}k_{l}^{2}/2m$  and $k_{l}$ is the wavenumber of the optical lattice lasers and $m$ is the atom mass.  The linear optical lattice is described by $-V\cos (x)$ with the lattice depth $V$. The nonlinear coefficient in the GP equation is
 \begin{equation}
 \mathcal{G}_\textrm{non}=g_{1}+g_{2} \cos (x). 
 \end{equation}
 The nonlinear lattice is described by $g_2\cos (x)$ with $g_2$ being the nonlinear-lattice amplitude. We also incorporate a constant interaction with the nonlinear coefficient $g_1$.  We consider that two lattices have the same spatial structure and the same period.  The relative phase between two lattices are controlled by the sign of $g_2$ and $V$. Concretely, we assume $g_2>0$ and change the sign of $V$ to analyze. When $V>0$ the two lattices are out of phase, and they are in phase when $V<0$. 

The experimental loading of BECs into the mixed nonlinear and linear lattices connects with Bloch states.  They are defined as $\psi(x,t)=e^{ikx-i\mu_k t}\phi_{k}(x)$. Here $k$ is the quasimomentum, $\mu_k$ is the chemical potential, and $\phi_{k}(x)$ is a periodic function having the same period as the mixed lattices, i.e.,  $\phi_{k}(x+2\pi)=\phi_{k}(x)$. Substituting the Bloch state solutions into the GP equation, we have, 
\begin{equation}
	\label{tgp}
	\mu_k\phi_k=-\frac{1}{2}(\frac{d}{d x}+ik )^2 \phi_{k}-V\cos (x) \phi_{k}+ \mathcal{G}_\textrm{non}|\phi_{k}|^{2}\phi_{k}.
\end{equation}
By solving above nonlinear equation with a normalization condition $\int_{0}^{2\pi} dx |\phi_k(x)|^2=1$, we can get Bloch spectrum $\mu(k)$ and the associated Bloch states $\phi_{k}$. In detail, we expand the periodic function $\phi_{k}$ using a plane-wave basis, and the above nonlinear equation turns to be coupled nonlinear ordinary equations for the plane-wave coefficients~\cite{Wu2003}, which can be solved using the standard Newton relaxation method.

Once we know the BEC Bloch states, we study their dynamical instability by linearizing the GP equation around the Bloch states.  We add perturbations to the Bloch states, 
\begin{equation}
	\label{perturbation}\begin{aligned}
	\psi (x,& t)	=e^{i k x-i\mu_k t} \\
&\times  [\phi_{k}(x)+u_{kq}(x) e^{i q x-i \omega_{kq} t}+v_{kq}^{*}(x) e^{-i q x+i \omega_{kq}^{*} t}],
\end{aligned}\end{equation}
where $q$ is the quasimomentum of perturbations, $\omega_{kq}$ is the energy of perturbations, and $u_{kq}(x)$ and $v_{kq}(x)$ are the perturbation amplitudes. After substituting the general wave function in Eq.~(\ref{perturbation}) into the GP equation and keeping only the linear terms relating to the perturbation amplitudes, we get 
the following Bogoliubov–de Gennes (BdG) equation,
\begin{equation}
\label{BdG}
	\omega_{kq}	\left(\begin{aligned}
		u_{kq}\\v_{kq}
	\end{aligned}\right)=\mathcal{H}_{\mathrm{BdG}}(k,q)	\left(\begin{aligned}
		u_{kq}\\v_{kq}
	\end{aligned}\right),
\end{equation}
where the BdG Hamiltonian is
\begin{equation}
\label{BdGHamiltonian}
	\mathcal{H}_{\mathrm{BdG}}(k,q)=	\begin{pmatrix}
	\mathcal{L}(k,q) & \mathcal{G}_\textrm{non}\phi_k^{2} \\
	-\mathcal{G}_\textrm{non}\phi_k^{*2}& 
	-\mathcal{L}(-k,q)
	\end{pmatrix},
\end{equation}
with
\begin{equation} \begin{aligned}
	\mathcal{L}&(k, q)\\
	&=-\frac{1}{2}\left[\frac{\partial}{\partial x}+i(k+q)\right]^{2}-V\cos(x)-\mu_k+2\mathcal{G}_\textrm{non}|\phi_k|^{2}.
\end{aligned}\end{equation}
The unique feature of the BdG Hamiltonian for a condensate is that it is non-Hermitian, i.e., $\mathcal{H}_{\mathrm{BdG}}^\dagger\ne \mathcal{H}_{\mathrm{BdG}}$.  Therefore, the BdG Hamiltonian allows for the existence of complex eigenvalues in $\omega_{kq}$. In the presence of complex modes in $\omega_{kq}$, it is known that the perturbation amplitudes in Eq.~(\ref{perturbation}) shall grow up exponentially with time, which means that a small perturbation shall deviate the evolution of the wave function far away from the condensed state. Consequently, the condensed state is dynamically unstable if there exists any complex mode in $\omega_{kq}$.  Through this dynamical instability, the condensed state is broken to lose superfluidity. We examine the dynamical instability of Bloch states by diagonalizing the BdG equation in Eq.~(\ref{BdG}).  We note that the BdG Hamiltonian is spatially periodic due to the Bloch states.  To carry out the diagonalization, we assume that the perturbation amplitudes are periodic functions which can be represented by a plane-wave expansion, so the BdG Hamiltonian is projected into the plane-wave basis and the resulted $\omega_{kq}$ are in the form of Bloch spectrum with the Brillouin zone $q\in (-0.5,0.5]$~\cite{He2021}. We are only interested in the Bloch states in the lowest band in Eq.~(\ref{tgp}).  Considering the Brillouin zone $k\in (-0.5,0.5] $ and the symmetry $\mu_k=\mu_{-k}$, we only analyze the instabilities of the Bloch states at $k\in [0,0.5]$ in the lowest band.  Meanwhile, the symmetry of the BdG Hamiltonian is
\begin{equation}
	\sigma_x\mathcal{H}_{\mathrm{BdG}}(k,q)\sigma_x=-\mathcal{H}^*_{\mathrm{BdG}}(k,-q),
\end{equation}
where $\sigma_x$ is a Pauli matrix.
With the symmetry, we know that if the eigenvalue of  $\mathcal{H}_{\mathrm{BdG}}$ is $\omega$ at $(k,q)$, then the eigenvalue immediately becomes $-\omega^*$ at  $(k,-q)$. Therefore, the perturbation energy has a symmetry $\omega_{kq}=-\omega^*_{k-q}$. Considering this symmetry, we only calculate the Bloch spectrum belonging to $q\in [0,0.5]$ in the BdG equation to check whether there exists any complex eigenvalue.  We define the growth rate $\Gamma$ to describe the instability. It is the maximum value of the imaginary parts of $\omega_{kq}$,
\begin{equation}
\label{GrowthRate}
\Gamma=\textrm{Max}[\textrm{Imag}(\omega_{kq})].
\end{equation}
 If the calculated growth rate is nonzero (zero), the corresponding Bloch state is dynamically unstable (stable). 

\begin{figure}[t]
	\centering	\includegraphics[width=1\linewidth]{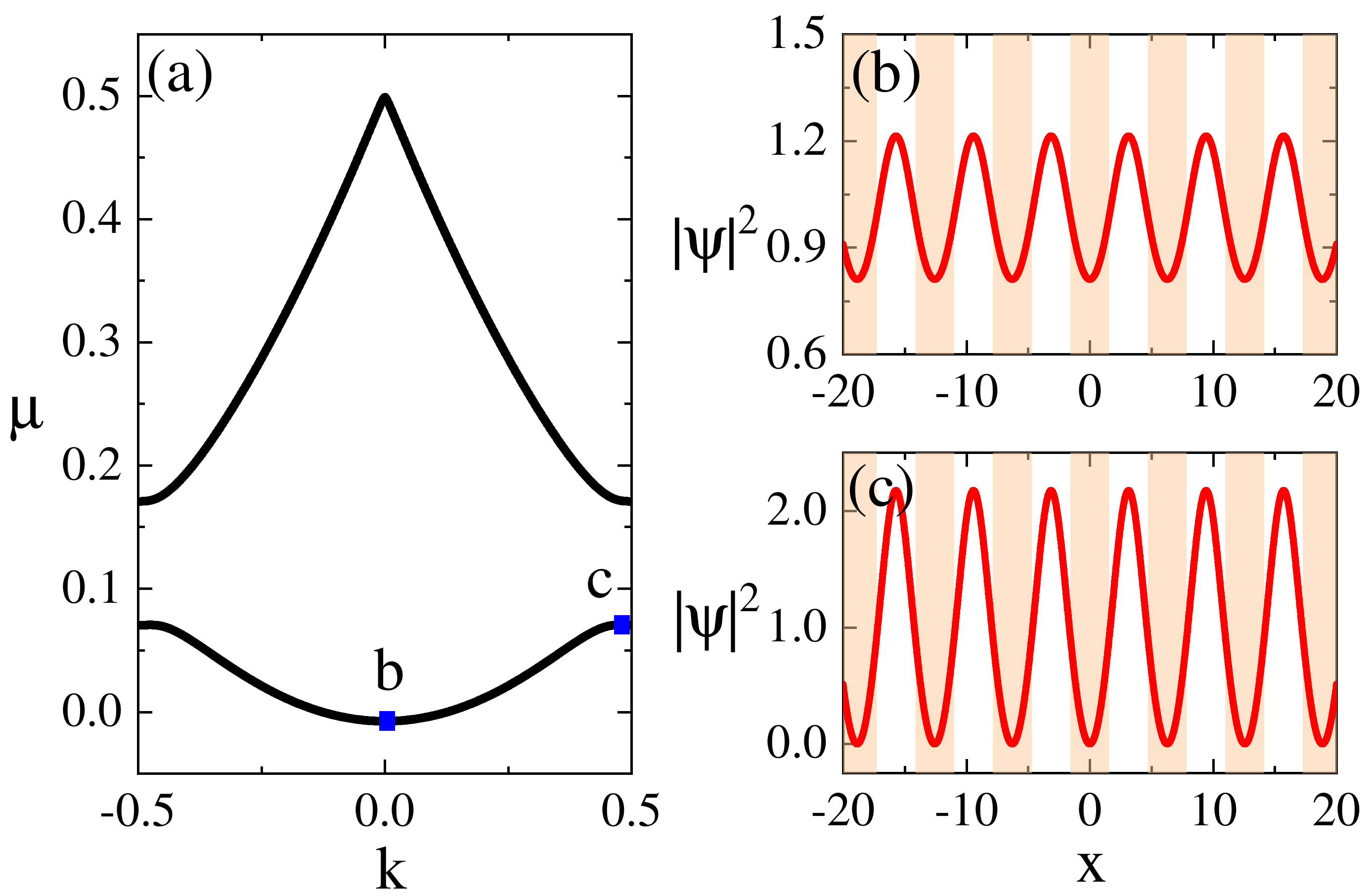}
	\caption{Nonlinear Bloch spectrum and associated nonlinear Bloch states of a BEC in a pure nonlinear lattice.  $g_{2}=0.05$ and $V=0$, $g_1=0$.
		(a) The lowest two Bloch bands. (b) and (c) The density distributions of  nonlinear Bloch states at Brillouin zone center and edge in the lowest band respectively [labeled by squares in (a)]. Pink stripes represent the regions that the nonlinear lattice is positive $g_{2}\cos x>0$.}
	\label{Fig1}
\end{figure}

We also examine the Landau instability of the Bloch states. In comparison with the linearization of the GP equation for the dynamical instability, the Landau instability needs to linearize  the energy functional of the system around the Bloch states~\cite{Wu2001}.  Small perturbations around the Bloch states generate an additional energy functional as~\cite{Wu2001}
\begin{equation}
\label{Landau}
\sigma_z\mathcal{H}_{\mathrm{BdG}}(k,q), \end{equation}
where $\sigma_z=diag(1,-1)$ is a Pauli matrix.
$\sigma_z\mathcal{H}_{\mathrm{BdG}}(k,q)$ is Hermitian so that its eigenvalues are real-valued. If there is any negative eigenvalue in 
$\sigma_z\mathcal{H}_{\mathrm{BdG}}(k,q)$, the corresponding Bloch states are not local minima of the energy functional and they are Landau unstable.  The occurrence of the Landau instability relates to Landau's criteria of superfluidity~\cite{Chen2010}. We use the same procedure as the treatment of the BdG Hamiltonian to diagonalize $\sigma_z\mathcal{H}_{\mathrm{BdG}}(k,q)$ to seek whether there exists any negative eigenvalue in $\sigma_z\mathcal{H}_{\mathrm{BdG}}(k,q)$.

\begin{figure}[t]
\centering
\includegraphics[width=0.956\linewidth]{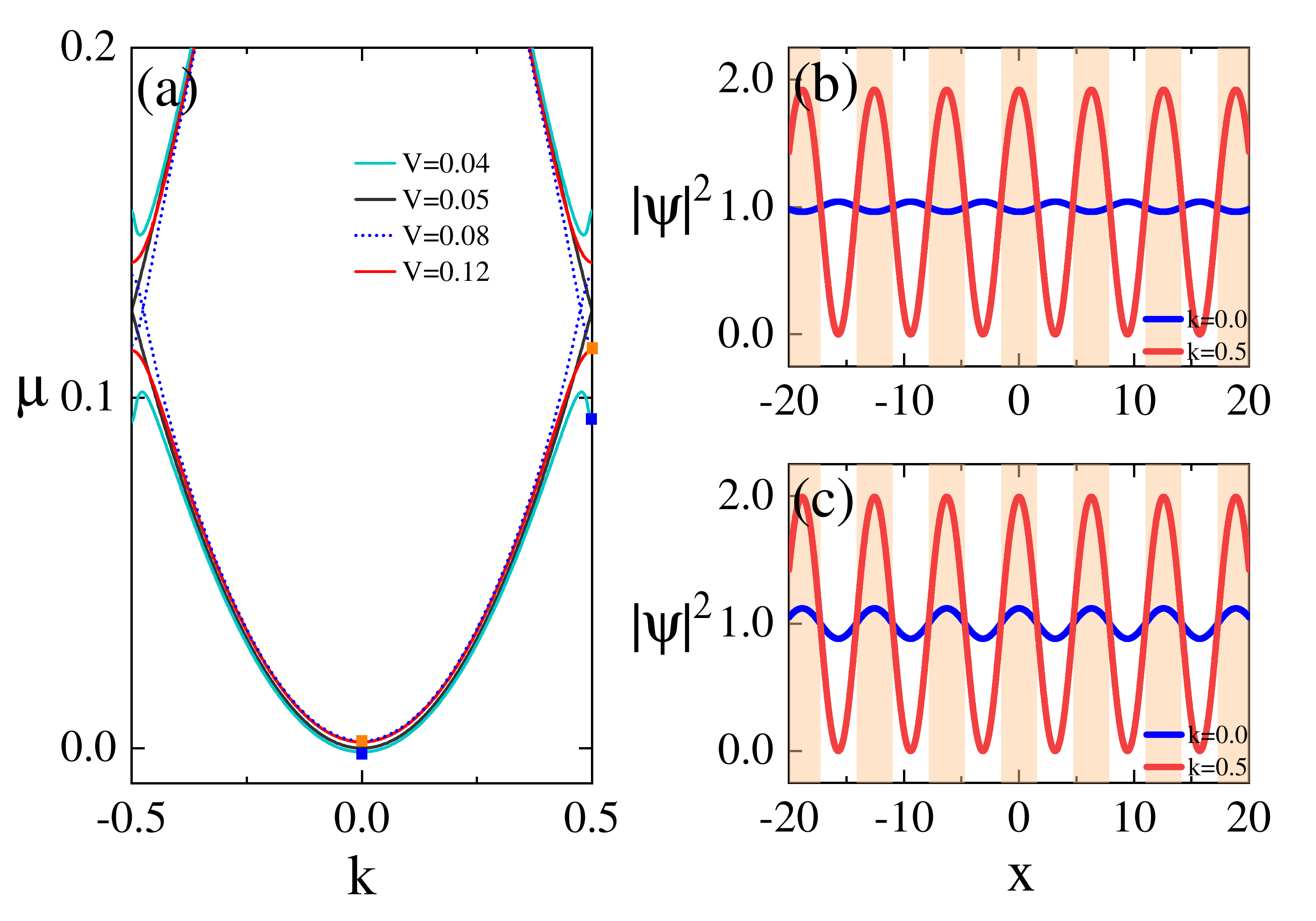}
\caption{Nonlinear Bloch spectrum and associated nonlinear Bloch states of a BEC with a nonlinear lattice and an out-of-phase linear lattice.  The nonlinear lattice amplitude is $g_{2}=0.05$ and the constant interaction  is $g_1=0$.
(a) The lowest two Bloch bands. When $V<0.05$ there is a gap opening between them (cyan-solid lines).   $V=0.05$ is a critical value where the lowest two bands connect at Brillouin zone edges (black-solid lines). When $V=0.08$ gap is still closed (dotted lines). Further increasing $V$ results in the gap reopening (red-solid lines). 
(b) and (c) The density distributions of  nonlinear Bloch states at Brillouin zone center (blue lines) and edge (red lines) in the lowest band for $V=0.04$ and $V=0.12$ respectively [labeled by squares in (a)]. Pink stripes represent the regions that the nonlinear lattice is positive $g_{2}\cos x>0$. Since the linear lattice is out-of-phase,  in the striped regions the linear lattice is negative $-V\cos x<0$.}
\label{Fig2}
\end{figure}
	
	\section{Nonlinear Bloch bands and associated Bloch states}
	\label{Bands}
	
We first study the existence of Bloch states in mixed nonlinear and linear lattices by solving Eq.~(\ref{tgp}). 	Fig.~\ref{Fig1} demonstrates nonlinear Bloch spectrum and associated Bloch states for a pure nonlinear lattice ($V=0$ and $g_1=0$). Only lowest two bands are shown in Fig.~\ref{Fig1}(a). It is interesting to find that the nonlinear Bloch spectrum is similar to that of a linear lattice. An energy gap is opened around Brillouin zone edges $k=\pm 0.5$ between the lowest two bands. The nonlinear Bloch state at $k=0$ in the lowest band is the lowest-energy state. Its density distribution is shown in Fig.~\ref{Fig1}(b).  The occupations in the minima of each nonlinear-lattice cell (white regions) dominates. Meanwhile, there are also populations in the maxima of the cells (shadowy regions) .  In contrast, the populations of the maxima for the Bloch state at the Brillouin zone edge in the lowest band are negligible [see Fig.~\ref{Fig1}(c)]. 

\begin{figure}[t]
\centering
\includegraphics[width=1\linewidth]{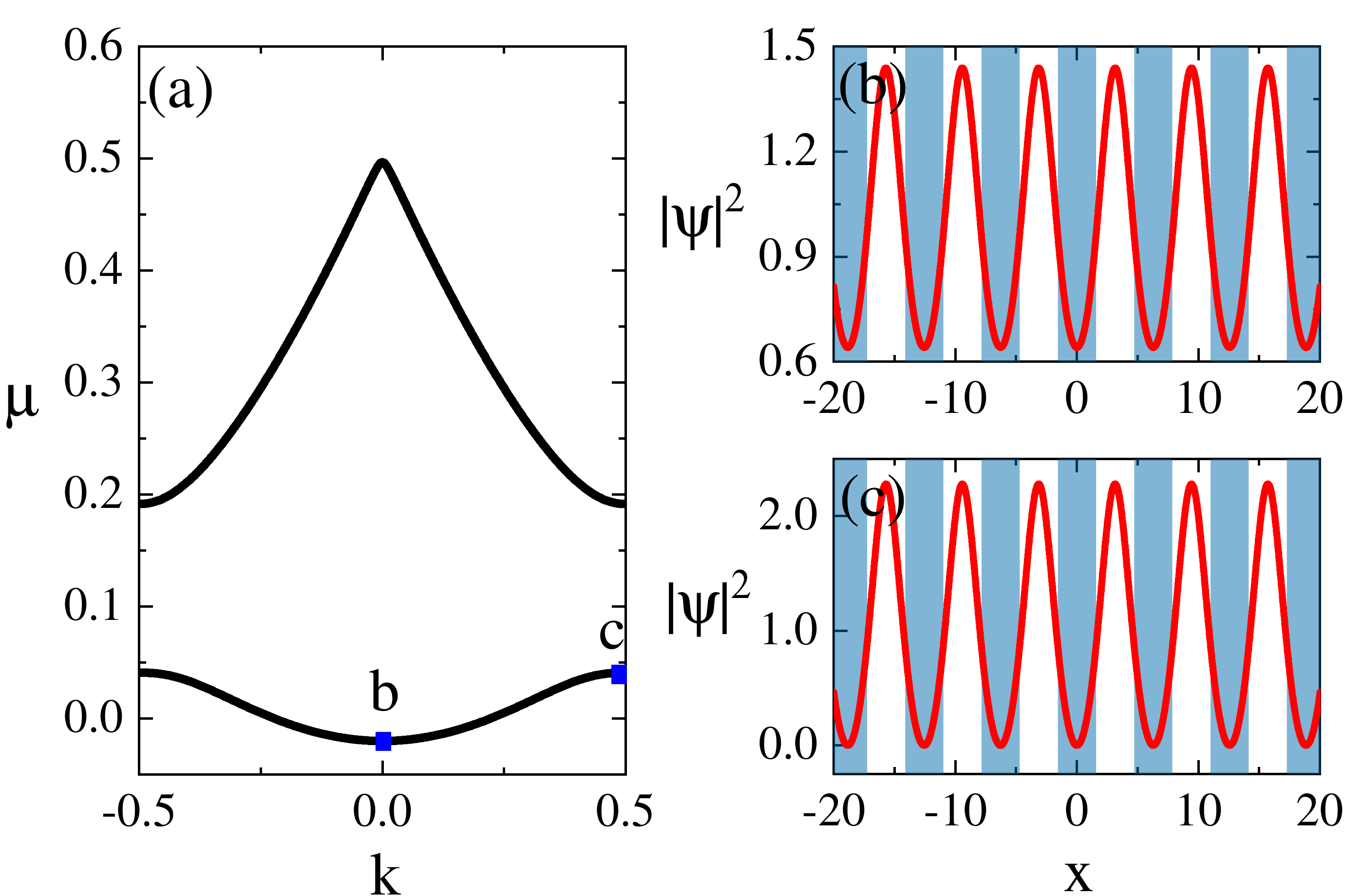}
\caption{Nonlinear Bloch spectrum and associated nonlinear Bloch states of a BEC with a nonlinear lattice and an in-phase linear lattice $V<0$.  $g_{2}=0.05$,  $V=-0.05$, and $g_1=0$. (a) The lowest two Bloch bands. (b) and (c) The density distributions of  nonlinear Bloch states at Brillouin zone center and edge in the lowest band respectively [labeled by squares in (a)]. Dark-blue stripes represent the regions that the nonlinear lattice is positive $g_{2}\cos x>0$, since the linear lattice is in-phase, it is also positive $-V\cos x>0$ in striped regions.}
\label{Fig3}
\end{figure}
	
Fig.~\ref{Fig2} demonstrates nonlinear Bloch bands with a nonlinear lattice and an out-of-phase linear lattice $ V>0$.  We find that $V=0.05$ is a critical value where the lowest two bands close gap and they connect at Brillouin zone edge (black-solid lines). Increasing $V$ from zero to the critical value, the size of the gap between them decreases.  Beyond the critical value, the gap is reopened (red-solid lines).  The gap between the lowest bands decreases to close and reopens is a signature of the competition between the two lattices. At the critical value $V=g_2$, the out-of-phase linear lattice completely cancels the effect of the nonlinear lattice.  When $V<g_2$ the nonlinear lattice dominates over the linear lattice. This can be seen from the density distribution of the Bloch state at $k=0$. As shown in Fig.~\ref{Fig2}(b), the occupations in the minima of nonlinear-lattice cells are still larger than these in the maxima [see the blue line]. This feature is the same as that in a pure nonlinear lattice. However, the Bloch state at the Brillouin zone edge chooses to occupy the cells of the linear lattice [see the red line in Fig.~\ref{Fig2}(b)].  When $V>g_2$ and the gap is reopened, the linear lattice surpasses the nonlinear one. Density distributes in the cells of the linear lattice for all Bloch states. Illustrating density distributions are demonstrated in Fig.~\ref{Fig2}(c) for $V=0.12$.

Fig.~\ref{Fig3} demonstrates nonlinear Bloch bands with a nonlinear lattice and an in-phase linear lattice. The linear lattice has the same phase as the nonlinear lattice. It enhances the effect of the nonlinear lattice.  Therefore, with the help of the linear lattice, the gap size between the lowest two bands is wider than that in a pure lattice [see Fig.~\ref{Fig3}(a)]. 
The Bloch states distribute inside the cells of both lattices [see Figs.~\ref{Fig3}(b) and~\ref{Fig3}(c)].

\begin{figure}[t]
\centering
\includegraphics[width=0.95\linewidth]{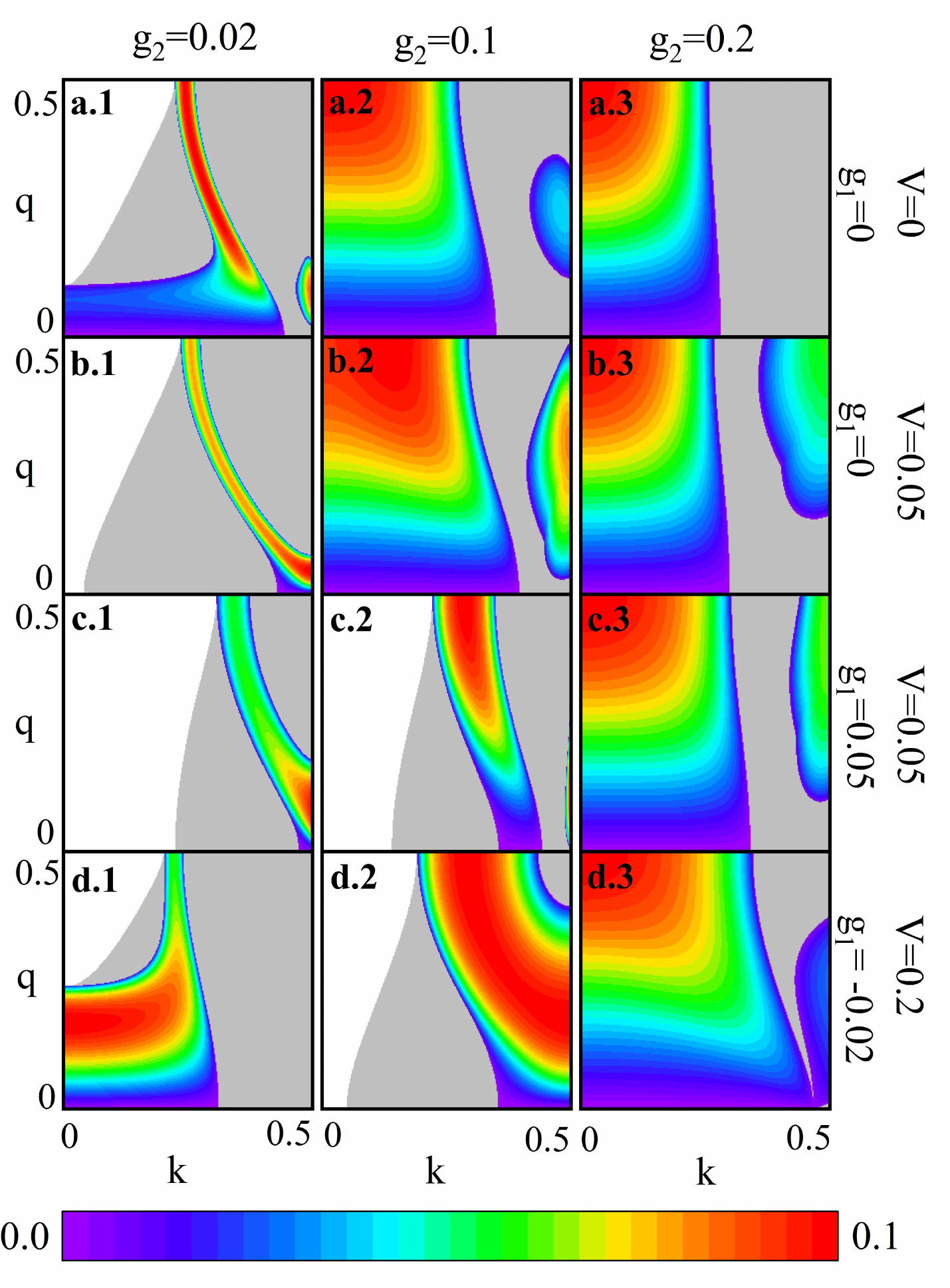}
\caption{Instabilities of the BEC Bloch states in the lowest Bloch band with a nonlinear lattice and an out-of-phase linear lattice $V>0$. $k$ and $q$ are the quasimomenta of the Bloch states and perturbations respectively.  The colored shadow areas represent that the Bloch states are dynamical unstable, and the color scale labels the growth rate $\Gamma$ defined in Eq.~(\ref{GrowthRate}); the scale changes from the dark purple $\Gamma=0$ to bright red $\Gamma=0.1$.  The gray areas indicate that the Bloch states have Landau instability. In the white regions, they are completely stable.  For a fixed Bloch state represented by a fixed $k$, if there is any unstable mode in a $q$, the corresponding Bloch state is unstable.  }
\label{Fig4}
\end{figure}

	\section{Instabilities of nonlinear Bloch states}
	\label{Instability}

\subsection{The out-of-phase linear lattices $V>0$}

The dynamical instability and Landau instability of the BEC Bloch states in the lowest Bloch band with mixed nonlinear and linear lattices are studied by diagonalizing the BdG Hamiltonian in Eq.~(\ref{BdGHamiltonian}) and the energy functional Hamiltonian in Eq.~(\ref{Landau}) respectively.	Fig.~\ref{Fig4} demonstrates typical results for out-of-phase linear lattices in the $(k,q)$ plane where $k$ and $q$ are the quasimomenta of the Bloch states and perturbations respectively. The results in the first row are for a pure nonlinear lattice with different amplitude $g_2$.  The pure nonlinear lattice has been studied in~\cite{Shaoliang2013, Dasgupta2016}. Our results are consistent with theirs.  In a pure nonlinear lattice, all Bloch states in the lowest band are Landau unstable (which represented  by gray areas in the plots). For a small amplitude $g_2$ (such as $g_2=0.02, 0.1$ in Figs.~\ref{Fig4}(a.1) and~\ref{Fig4}(a.2)),  the Bloch states at a finite region of $k$ close to Brillouin zone edge are dynamically stable (represented by the out of the colored areas).  When the amplitude is $g_2=0.2$ in Fig.~\ref{Fig4}(a.3) the Bloch states around Brillouin zone edge are dynamically stable.  The outstanding feature for the pure nonlinear lattice is that the Bloch states around Brillouin zone center $k=0$ are dynamically unstable. 	
	
In the presence of an out-of-phase linear lattice ($V=0.05$ in plots in the second row) the instabilities of the Bloch states change dramatically for a small $g_2$. 	Fig.~\ref{Fig4}(b.1) shows that the Bloch states around $k=0$ become both dynamically and Landau stable. Especially, the dynamically unstable Bloch states shrink to $k\in [0.25,0.5]$, which means that the Bloch states at $k\in [0,0.25)$ are dynamically stable. The results indicate that an out-of-phase linear lattice can stabilize the lowest-energy Bloch states against the dynamical and Landau instabilities.  The stabilization only works when the linear lattice dominates over the nonlinear lattice, i.e., $V>g_2$. If the nonlinear lattice dominates, the instabilities are similar to these in a pure nonlinear lattice, and the typical examples are shown in Figs.~\ref{Fig4}(b.2) and~\ref{Fig4}(b.3). 

\begin{figure}[t]
\centering
\includegraphics[width=0.95\linewidth]{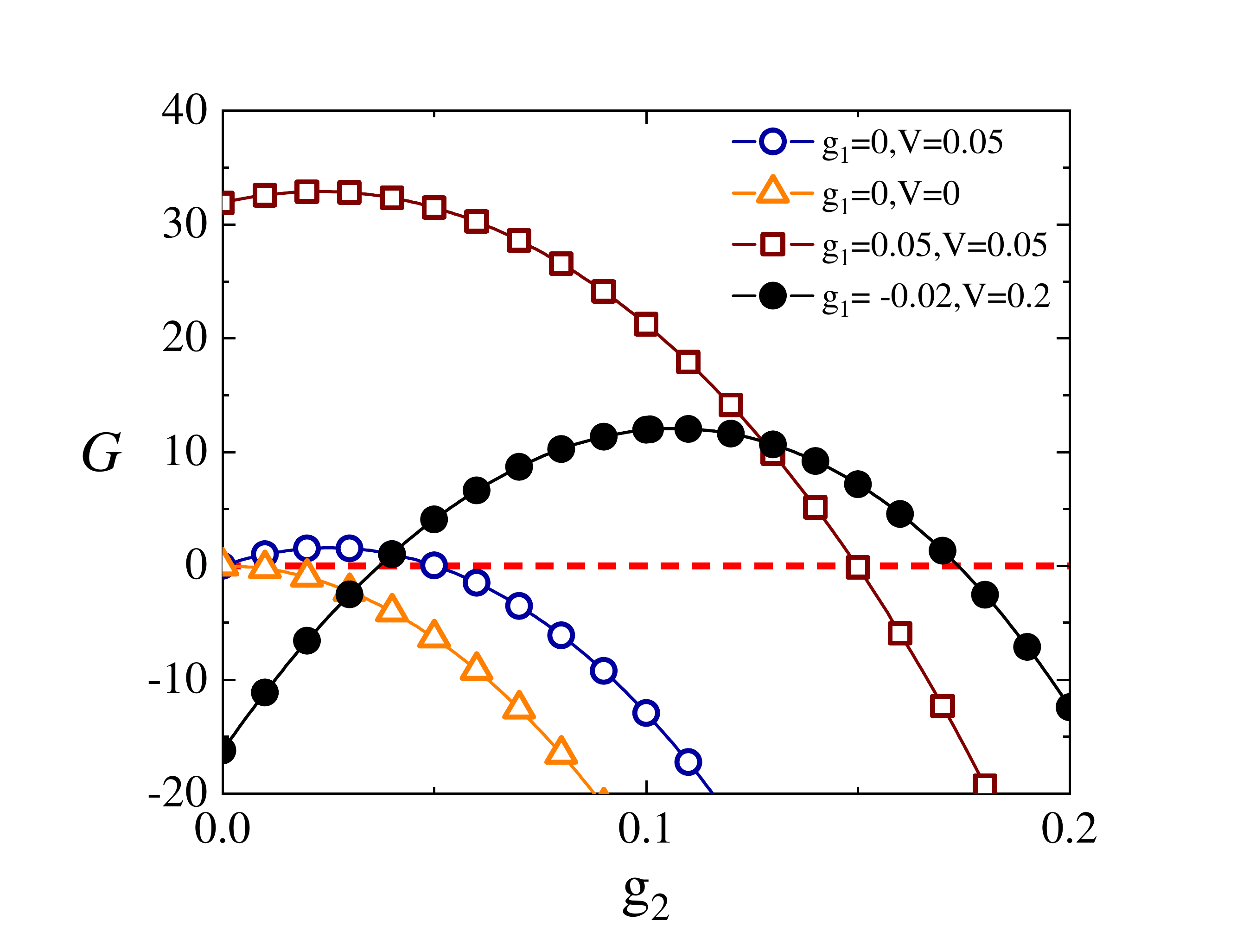}
\caption{The averaged interaction $G$ 	(defined in Eq.~(\ref{Averge})) of the Bloch state at $k=0$ with a nonlinear lattice and an out-of-phase linear lattice. 	The horizontal red dashed line is $G=0$ for guiding eyes. }
\label{Fig5}
\end{figure}

We use the averaged interaction firstly introduced in Ref.~\cite{Shaoliang2013} to uncover the mechanism of the stabilization of the lowest-energy Bloch state by the out-of-phase linear lattice.    The averaged interaction $G$ is defined as
\begin{equation}
\label{Averge}
G=\int_0^{2 \pi} d x \left[ g_1+g_2 \cos (x) \right] |\phi_k|^4.
\end{equation}
It represents the average value of the nonlinear energy over a period.  In Fig.~\ref{Fig5}, we plot the averaged interaction of the $k=0$ Bloch state as a function of $g_2$.  The blue-circle line is for the parameters of $V=0.05$ and $g_1=0$, corresponding to the second row in Fig.~\ref{Fig4}. It shows that the averaged interaction is repulsive (i.e., $G>0$) when $0<g_2<0.05$ and is attractive if $g_2>0.05$. It is known that  the Bloch state at $k=0$ in a linear lattice  is only dynamically and Landau stable when a constant nonlinearity is repulsive~\cite{Wu2001}. The averaged interaction behaves as an effective nonlinearity that the condensed atoms feel.  If it is repulsive, it is reasonable that the $k=0$ Bloch state is stable in the presence of the linear lattice. Oppositely, an attractive averaged interaction can not stabilize the $k=0$ Bloch state.  In the absence of the linear lattice, the calculated averaged interaction of the $k=0$ Bloch state is shown by the orange-triangular line in Fig.~\ref{Fig5}.  All of them are attractive, which results in the dynamical and Landau instabilities, and this expectation is consistent with the results demonstrating in the first row in Fig.~\ref{Fig4}. In the presence of a dominating out-of-phase linear lattice,  it changes density distributions of the Bloch state so that the averaged interaction may become repulsive.  Therefore, the out-of-phase linear lattice provides an experimentally accessible means to stabilize the lowest-energy Bloch state.

\begin{figure}[t]
\centering
\includegraphics[width=1\linewidth]{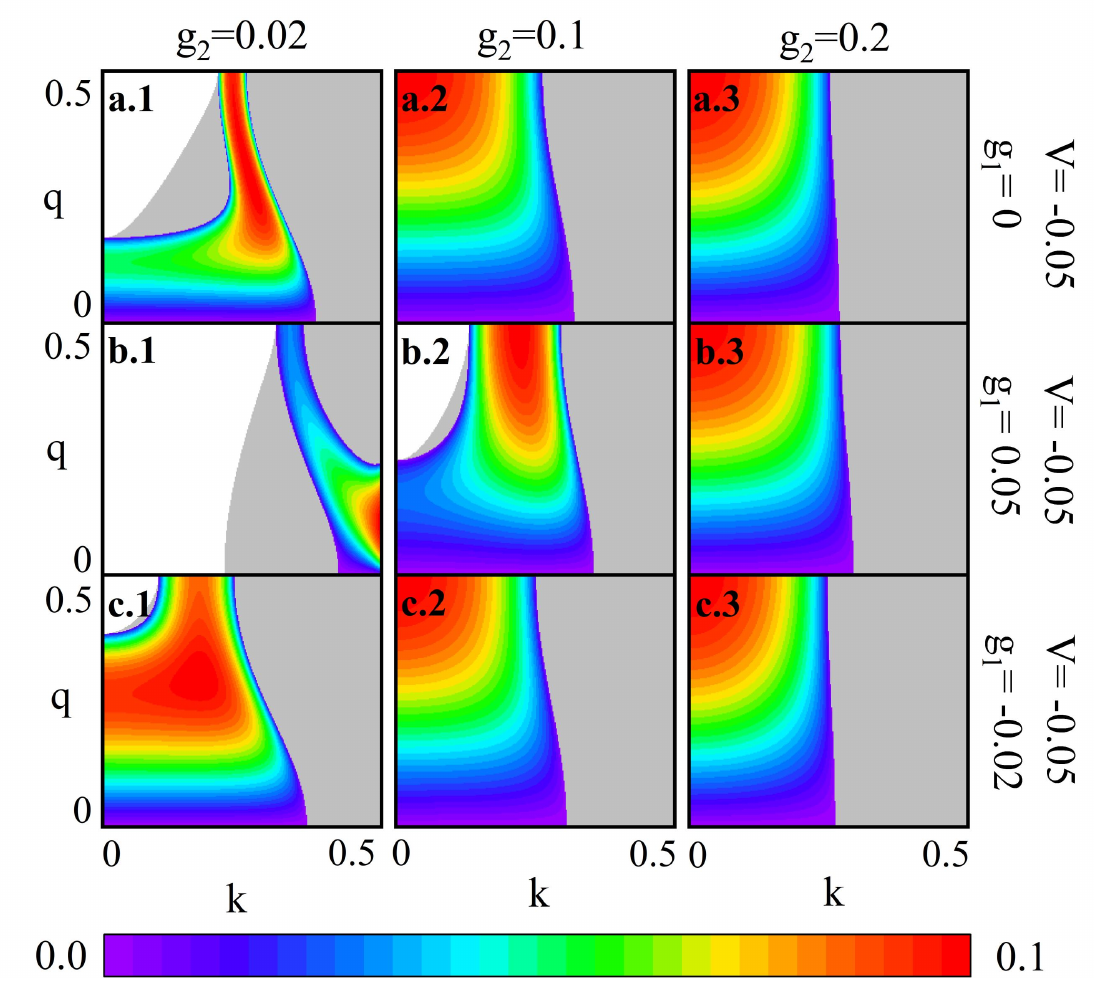}
\caption{
Instabilities of the BEC Bloch states in the lowest Bloch band with a nonlinear lattice and an in-phase linear lattice $V<0$. The colored shadow areas represent that the Bloch states are dynamical unstable, and the color scale labels the growth rate $\Gamma$ defined in Eq.~(\ref{GrowthRate}); the scale changes from the dark purple $\Gamma=0$ to bright red $\Gamma=0.1$.  The gray areas indicate that the Bloch states have Landau instability. In the white regions, they are completely stable. }
\label{Fig6}
\end{figure}

We also incorporate a nonzero constant interaction $g_1\ne 0$ into the out-of-phase linear lattice. The results are shown in the third row in Fig.~\ref{Fig4} for  a repulsive  constant interaction $g_1=0.05$ and $V=0.05$.   For the small nonlinear-lattice amplitude, such as $g_2=0.02$ and   $g_2=0.1$ in Figs.~\ref{Fig4}(c.1) and~\ref{Fig4}(c.2),  the stable regions around $k=0$ (white areas) become very wide.  In comparison with the results of $g_1=0$ in the second row, the repulsive constant interaction $g_1$ further enhances the stability of the Bloch states around $k=0$. However, if $g_2$ dominates, such as $g_2=0.2$ in Fig.~\ref{Fig4}(c.3), the structure of instabilities becomes the same with a pure nonlinear lattice.  The brown-square line in Fig.~\ref{Fig5} describes the averaged interaction of the $k=0$ Bloch states for this case. It shows that up to a critical $g_2$ the averaged interaction is repulsive and beyond the critical $g_2$ it becomes attractive due to the dominating $g_2$.  The repulsive constant interaction broadens the repulsive area of the averaged interaction comparing with the case of $g_1=0$ and $V=0.05$. This is because that $g_1>0$ itself contributes repulsively to $G$ in Eq.~(\ref{Averge}). So a repulsive constant interaction is favorable for the stabilization of the $k=0$ Bloch state with mixed lattices.   $g_1<0$ contributes attractively to $G$, so the stabilization can not be benefited from $g_1<0$.  Surprisingly,  We still find the repulsive averaged interactions with a large linear lattice $V=0.2$  and an attractive constant $g_1=-0.02$. The result is shown by the black-dot line in Fig.~\ref{Fig5}. In the middle region of $g_2$, the averaged interaction is repulsive. The instability results demonstrated in the forth row in Fig.~\ref{Fig4} confirm that a small $g_2$ in~\ref{Fig4}(d.1) and a large one in~\ref{Fig4}(d.3) lead to instabilities to the Bloch states around $k=0$ and the middle value as $g_2=0.1$ in ~\ref{Fig4}(d.1) results in a stable Bloch state at $k=0$.

\subsection{The in-phase linear lattices $V<0$}

The typical results of the BEC Bloch states with a nonlinear lattice and an in-phase linear lattice are described in Fig.~\ref{Fig6}. The first row shows the results of the in-phase linear lattice $V=-0.05$ and $g_1=0$. All Bloch states are Landau unstable. Outstandingly, the states around Brillouin zone edge $k=0.5$ (center $k=0$) are dynamically stable (unstable). The physical reason is that the in-phase linear lattice enhances the effect of the nonlinear lattice since the structures of two lattices are spatially matched.  Therefore, the averaged interactions $G$ for the states at $k=0$ and at $k=0.5$ are always attractive.  Ref.~\cite{Barontini2007} have revealed that the Bloch states with attractive interactions in a linear lattice are always Landau unstable and  are dynamically stable (unstable) around the Brillouin zone edges (center). So it is the attractive averaged interaction that makes the states around $k=0$ ($k=0.5$) dynamically unstable (stable).

\begin{figure*}[t]
\centering
\includegraphics[width=1\linewidth]{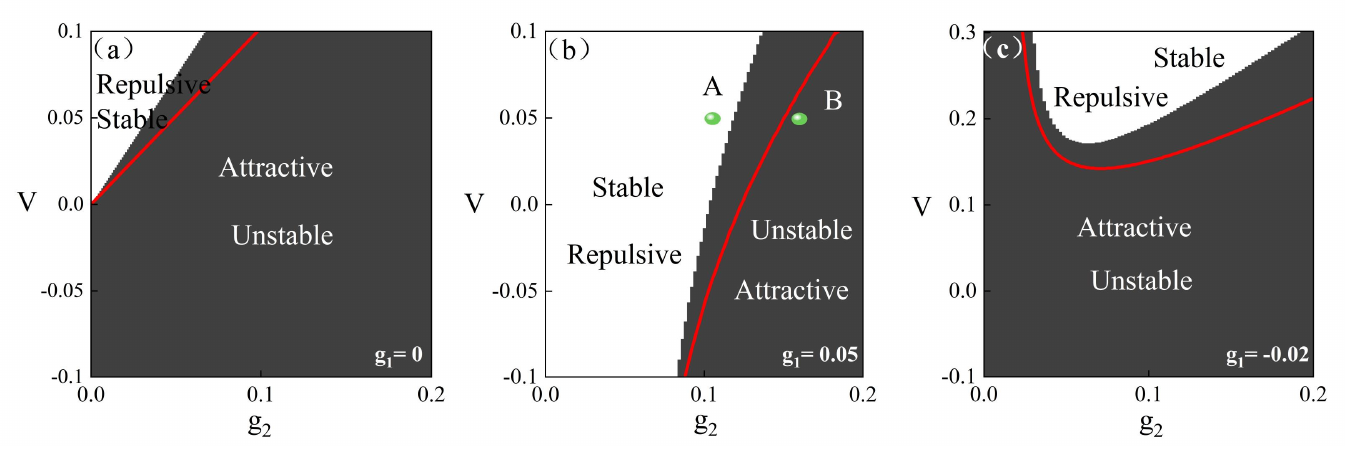}
\caption{The  dynamical-instability-phase-diagram of the BEC Bloch states at Brillouin zone center $k = 0$ with the mixed nonlinear and linear lattices in the parameter space $(g_2, V)$.  (a) $g_{1}=0$, (b) $g_{1}=0.05$, and (c) $g_{1}= -0.02$.  In the white regions, the $k=0$ Bloch  state is dynamically stable; in the dark regions, the Bloch state is dynamically unstable. The red lines represent the zero averaged interaction $G=0$, and in the regions above the red lines the averaged interaction is repulsive and in the other regions it is attractive.}
\label{Fig7}
\end{figure*}
\begin{figure*}[ht]
\centering
\includegraphics[width=1\linewidth]{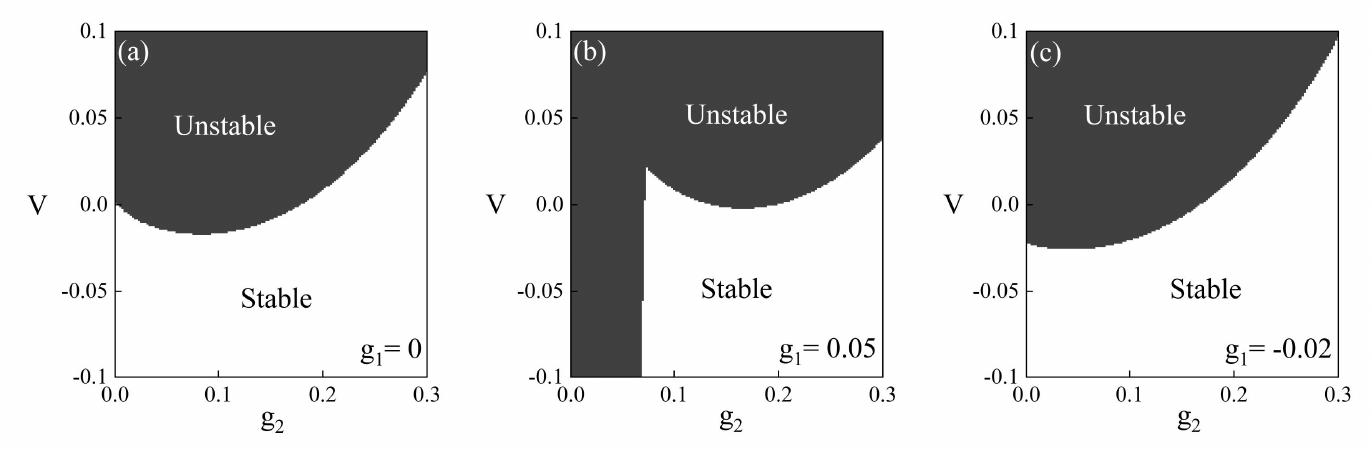}
\caption{The  dynamical-instability-phase-diagram of the BEC Bloch states at Brillouin zone edge $k = 0.5$ with the mixed nonlinear and linear lattices in the parameter space $(g_2, V)$.  (a) $g_{1}=0$, (b) $g_{1}=0.05$, and (c) $g_{1}= -0.02$. In the white regions, the $k=0.5$ Bloch  state is dynamically stable; in the dark regions, the Bloch state is dynamically unstable. }
\label{Fig8}
\end{figure*}

We also add a constant interaction $g_1$ into the mixed lattices.  The second row in Fig.~\ref{Fig6} is the results for a repulsive interaction $g_1=0.05$, and the third row is these for an attractive one $g_1=-0.02$.  The repulsive constant interaction in Fig.~\ref{Fig6}(b.1) is dominant, therefore, the instability structures in the ($k,q$) plane are similar to these of a BEC with repulsive interactions in a linear lattice~\cite{Wu2001}. When the repulsive constant interaction losses the dominant role, the instabilities become the same as these with only the mixed lattices [see Figs.~\ref{Fig6}(b.2) and ~\ref{Fig6}(b.3)].  On the other hand, an attractive  constant interaction has the same effect with the mixed lattice.  The third row shows that the presence of an attractive constant interaction $g_1=-0.02$ does not qualitatively modify the instability structures in comparison with the first row.

\subsection{Dynamical instabilities of Bloch states at Brillouin zone center and edge}

The BEC experiment has shown that the trigger of the Landau instability requires a long time and the dynamical instability happens in a short time~\cite{Fallani2004}. Therefore, the dynamical instability may be more relevant in experiments. Furthermore, the Bloch states at Brillouin zone center and edges are distinctively interesting due to their high symmetries.  Here, we summarize their dynamical instabilities studied in previous sections to clearly show that the linear lattice can be an efficient approach to stabilize unstable Bloch states of a nonlinear lattice. 
	
Fig.~\ref{Fig7} is the dynamical-instability-phase-diagram of the $k=0$ Bloch states in the space of $(g_2,V)$. 
Fig.~\ref{Fig7}(a)  is the case of a zero constant interaction, $g_1=0$.  The white area represents that the state is dynamically stable.  Only the out-of-phase linear lattice $V>0$ could stabilize it. The red line corresponds to zero averaged interaction of the $k=0$ Bloch state, $G=0$, above which $G$ is repulsive.  Note that the boundary between the white (stable) and dark (unstable) areas is slightly mismatched with the line $G=0$.  This means that the mechanism to stabilize the $k=0$ Bloch state by the linear-lattice-induced repulsive averaged interaction is not exact. However, the mechanism truly provides an intuitive and qualitative way for our understanding of the stabilization.  With the help of a repulsive constant interaction $g_1=0.05$ in Fig.~\ref{Fig7}(b),  the stabilization is extended from the out-of-phase lattice $V>0$ to the in-phase one $V<0$.  Even the constant interaction is attractive, such as $g_1=-0.02$ in Fig.~\ref{Fig7}(c),  we still find that a large out-of-phase lattice can stabilize the states with a finite $g_2$.

Fig.~\ref{Fig8} is the dynamical-instability-phase-diagram of the $k=0.5$ Bloch states in the space of $(g_2,V)$. 
In the absence of  the constant interaction $g_1=0$ in Fig.~\ref{Fig8}(a),  the effect of the linear lattice reflects two aspects: the in-phase lattice $V<0$ always strengths the stability of the $k=0.5$ state; the out-of-phase lattice $V>0$ weakens its stability in the sense that the lattice increases the critical value of $g_2$ beyond which the state becomes stable. In the presence of a repulsive constant interaction $g_1=0.05$ in  Fig.~\ref{Fig8}(b),  it is dominant if $g_2$ and $-V$ are small, which destabilizes the state. However, for an attractive constant interaction $g_1=-0.02$ in Fig.~\ref{Fig8}(c), the diagram is qualitatively same with the $g_1=0$ case in Fig.~\ref{Fig8}(a).

\begin{figure}[t]
\centering
\includegraphics[width=1\linewidth]{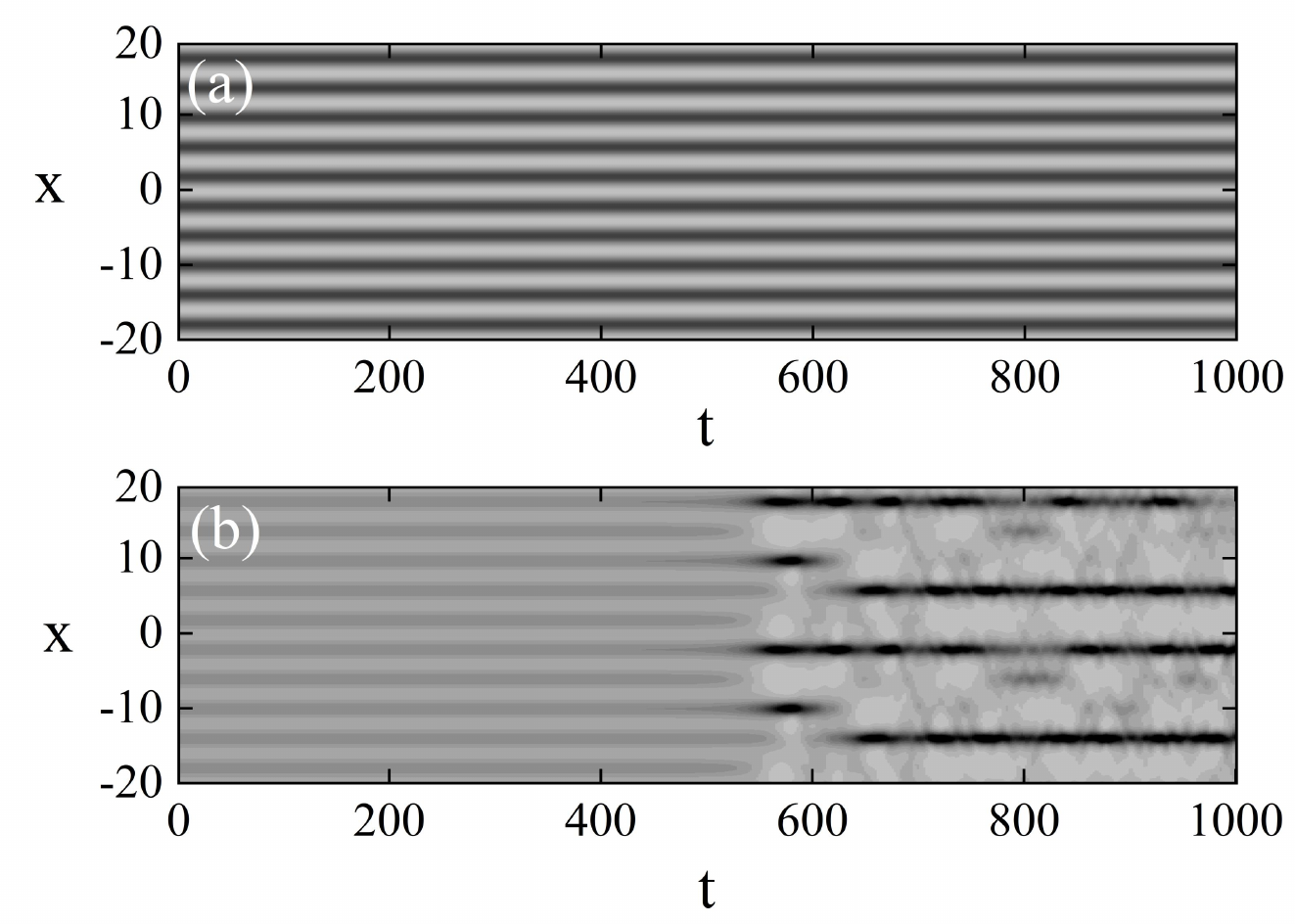}
\caption{The time evolution of the $k=0$ Bloch states represented by the marked points in Fig.\ref{Fig7}(b) with $g_{1}=0.05$.  Plots show the density distributions as a function of time $t$. (a) A stable evolution $g_{2}=0.1$ and $ V=0.1$. (b) An unstable evolution $g_{2}=0.15$ and $V=0.1$.}
\label{Fig9}
\end{figure}		

Finally, we comment that the dynamical instability of the Bloch states calculated from the BdG equation in  Eq.~(\ref{BdG}) can also be examined by the direct evolution of the GP equation in Eq.~(\ref{gp}) with the corresponding Bloch states serving as initial states.   Fig.~\ref{Fig9} shows typical examples of evolution.  The two $k=0$ Bloch states are represented by the marked points in Fig.~\ref{Fig7}.  The one of them is known to be dynamically stable and the other is unstable from the calculation of the BdG equation. We set them as initial states to evolve the GP equation.  As expected, the stable state evolves stably [see Fig.~\ref{Fig9}(a)] and the unstable state breaks down during the evolution [see Fig.~\ref{Fig9}(b)].  The time evolution of the Bloch states offers an experimental approach to examine the instability.   In the experiment~\cite{Yamazaki2010}, the mixed nonlinear and linear lattices with the same period can be implemented by the optical Feshbach resonance of an optical standing wave.  Following this experiment, we propose to load the BEC into the $k=0$ Bloch state by adiabatically ramping up the standing wave. The system is then held for a certain time to let free evolution of the Bloch state. Finally, the decay of condensed atom number is observed, from which the loss rate is measured. The loss rate is relevant to the growth rate defined in Eq.~(\ref{GrowthRate}).

\section{Conclusion}
\label{Conclusion}

BECs in periodic potentials give rise to interesting physics relevant to instabilities of Bloch states.  Their instabilities are experimentally involved to relate to the breakdown of BEC superfluidity. It has been shown that even the lowest-energy Bloch state is unstable for the BECs in a nonlinear lattice which challenges its experimental implementations. We propose to add a linear lattice to the BECs with the nonlinear lattice  to stabilize the lowest-energy Bloch state. We systematically study the instabilities of BEC Bloch states in mixed nonlinear and linear lattices. The two lattices have the same spatial structure and the same period, but leaving the relative phase is tunable. We find that an out-of-phase linear lattice enables to make Bloch states around Brillouin zone center to be dynamically and Landau stable. The stabilization mechanism is revealed as the out-of-phase lattice changes density distributions to induce repulsive averaged interactions. In contrast, an in-phase linear lattice enhances the effect of the nonlinear lattice and can not change density distributions. It always induces attractive averaged interactions, therefore it is useless for the stabilization.  It is known that Bloch states around Brillouin zone edge  become dynamically stable in the pure nonlinear lattice when the lattice amplitude is beyond a critical value. The presence of the out-of-phase lattice moves the critical value to be more large and the in-phase lattice assists to make them dynamically stable no matter the value of the nonlinear-lattice amplitude. 

We also incorporate a constant interaction into the BECs with mixed nonlinear and linear lattices.  A repulsive constant interaction extends the out-of-phase-linear-lattice-induced stabilization of the Bloch states around Brillouin zone center to the in-phase linear lattice. Even in the presence of an attractive constant interaction,  we find the out-of-phase linear lattice still can stabilize the states.  For the Bloch states around Brillouin zone edges, the constant interaction, no matter attractive or repulsive, does not qualitatively change their instability properties.

\section{Acknowledges}
This work was supported by National Natural Science Foundation of China with Grants No.11974235 and 11774219.

\end{document}